\documentstyle[prb,eqsecnum,aps,epsf,multicol]{revtex}

\begin{document}
\draft

\title{Mode locking in Quantum Hall effect point contacts }
\author{Hsiu-Hau Lin}
\address{Department of Physics, University of California\\
Santa Barbara, CA 93106-9530}

\author{Matthew P. A. Fisher}
\address{Institute for Theoretical Physics, University of California\\
Santa Barbara, CA 93106-4030} 
\date{\today} 
\maketitle

\begin{abstract}
We study the effect of an ac drive on the current-voltage (I-V)
characteristics of a tunnel junction between two
fractional Quantum Hall fluids
at filling $\nu ^{-1}$ an odd integer.  
Within the chiral Luttinger liquid model
of edge states, the point contact dynamics is described by a driven
damped quantum mechanical pendulum.  In a semi-classical limit
which ignores electron tunnelling,
this model exhibits mode-locking, which corresponds
to current plateaus in the I-V curve at integer multiples
of $I= e\omega /2\pi$, with $\omega$ the ac drive angular frequency.
By analyzing the full quantum model at non-zero $\nu$
using perturbative and exact methods,
we study the effect of quantum fluctuation on the mode-locked
plateaus.  For $\nu=1$ quantum fluctuations smear completely the plateaus,
leaving no trace of the ac drive.  For $\nu \ge 1/2$ smeared plateaus
remain in the I-V curve, but are not centered at the currents
$I=n e \omega /2\pi$.  For $\nu < 1/2$ rounded plateaus centered
around the quantized current values are found.  The possibility of using
mode locking in FQHE point contacts as a current-to-frequency standard
is discussed.  
\end{abstract}


\begin{multicols}{2}

\section{Introduction}
The conductance through a tunnel junction 
is proportional to the electron density of states in the two electrodes.
For metallic electrodes,
which have a non-zero density of states 
at the Fermi energy, the tunnel junction current-voltage (I-V) characteristics
are Ohmic at low bias.  
In marked contrast, recent theories\cite{wen,moon} have predicted
strongly non-ohmic behavior for 
tunneling through a point contact separating
two fractional quantum Hall effect (FQHE) fluids.  
Specifically, for filling
factor $\nu =1/m$, with odd integer $m$,
the tunnel current at zero temperature is predicted to vary with voltage
as $I \sim V^{2/ \nu -1}$.
At finite temperatures, Ohmic behavior is
recovered at small voltages, with 
a zero bias differential conductance varying as,
$dI/dV \sim T^{2/ \nu -2}$.  A temperature dependence
consistent with this has been seen
in a recent experiment by Milliken et. al.\cite{webb} for the  
tunnelling conductance
between two FQHE fluids at filling $\nu=1/3$.

The non-Ohmic tunneling conductance is due to
the strange properties of the edge states in the FQHE.  
FQHE edge states are a beautiful realization
of one-dimensional Luttinger liquids\cite{wen}. 
In contrast to metallic electrodes,  
the tunneling density of states in a Luttinger liquid 
{\it vanishes} at the Fermi energy, which leads to the vanishing
tunnel conductance between two FQHE fluids.
Thus, in contrast to conventional metallic tunnel junctions,
a FQHE tunnel junction is an {\it insulator}.

An {\it insulating} point contact junction is, in many respects,
the dual of a superconducting point contact - namely a
Josephson junction.  In a Josephson junction
the I-V curve is also strongly non-Ohmic, with
voltage vanishing rapidly for
currents below the critical current, $I_J$.
Moreover, the zero bias resistance
is expected to vanish exponentially as $T \rightarrow$,
$dV/dI \sim exp(-E_J/k_BT)$, with energy barrier $E_J = \phi I_J$.
Under exchange of current with voltage, the behavior is similar
to the vanishing conductance in the FQHE
point contact.  In a Josephson junction, the phase difference
between the superconducting electrodes is behaving classically,
whereas in the FQHE junction the classical variable is the transferred
electron charge.

One of the most striking manifestations of the ac Josephson effect,
is the presence of quantized voltage steps (Shapiro steps) in an applied
microwave field\cite{shapiro}.  The applied radiation at 
angular frequency $\omega$
mode locks to the discrete phase slip events leading to
plateaus at voltages $V = n(\hbar /2e)\omega$,
for integer $n$.  In the plateaus, the voltage
is so accurately quantized that Shapiro steps serve as a voltage-to-frequency
standard.  

The duality between Josephson junctions
and FQHE junctions, suggests that the latter might
also exhibit interesting behavior in the presence of an applied
ac field.  In this paper, we study in detail the effect of an ac
drive on a FQHE tunnel junction, focussing on the structure
induced in the $I-V$ characteristics.
One anticipates the possibility of mode locking between
the ac drive and the electron tunnelling events.
This could lead to steps in the junction {\it current}, 
quantized at integer multiples of $I=e \omega /2 \pi$ - 
the analog of Shapiro steps.

Quantized current plateaus for metallic tunnel junctions
were proposed several
years back\cite{ben-jacob}.
Due to Coulomb blockade effects,
it was argued that normal metal tunnel junctions with sufficiently
high resistances would exhibit the phenomena of Bloch oscillations - 
an oscillatory voltage in the presence of a dc current - the dual
of the ac Josephson effect.
Moreover, it was suggested that
an applied ac drive would mode lock
to these oscillations resulting in current plateaus.
A more favorable geometry
for current plateaus, consists of multiple tunnel junctions in
series, which can be separately tweaked by an ac drive,
thereby transferring the electrons one-by-one
through the circuit.
Such an electron ``turnstile" was realized experimentally,
by a number of groups, both in metallic systems\cite{geerligs,geerligs1,white}
and in 
semiconductor heterostructures\cite{kouwenhoven,kouwenhoven1,nagamune}.  
Due to the multiple
junction geometry the tunstiles only work well at
rather low frequencies, below tens of Megahertz.  At higher
frequencies the electrons take too long to pass
across the junctions, and do not ``keep up" with the ac drive.

In a Josephson junction Shapiro steps are very robust,
and do not need complicated multiple junction geometries.
Moreover, Shapiro steps are observed up to quite frequencies,
comparable to the superconducting gap.
The reason for this is that
the junction phase difference is a classical field,
so that phase slip processes are classical events which
readily lock to an ac drive.
In ``insulating" FQHE point contacts 
the electron charge is a good quantum number, which suggests that
mode locking might also be possible
in a single junction configuration.
However, quantum fluctuations in the electron charge transfer
are expected to be more important than quantum phase slip processes
in the Josephson junction, as reflected in the power law
voltage and temperature dependences in the I-V curves of the FQHE junction.
(Because the phase of the superconducting wave function exhibits
true long-ranged order, low frequency quantum phase slips
are expected to be completely absent.)
This paper is devoted to studying the effect of such quantum fluctuations
in washing out mode-locked steps.

The organization and central results of the paper are as follows.
In Section II we introduce the edge state model for
a FQHE tunnel junction at filling $\nu=1/m$, in the presence of both 
a dc source-to-drain voltage, $V_{sd}$, and an
a.c drive voltage, $V_{ac}\, \sin\omega t$.  While the model
is only appropriate for FQHE edges when $\nu^{-1}$ is an odd integer,
it is well defined for general $\nu$.    

In Section III we consider a semi-classical limit, 
which ignores quantum tunnelling
of the electron.  In this limit, the model reduces to the
classical dynamics of a periodically driven overdamped pendulum, with the phase
of the pendulum representing the charge transferred
across the junction.  This classical model is equivalent to the 
resistively-shunted junction (RSJ) model
of Josephson junction dynamics\cite{mccumber,stewart,johnson}.
Not surprisingly,
robust mode locked current plateaus are found in this semi-classical limit.

In Section IV we study the full quantum model,
and derive exact solutions for the I-V curves at two special values,
$\nu=1$ and $\nu=1/2$.  At $\nu=1$, appropriate for the integer
quantum Hall effect, quantum fluctuations are so strong that
{\it all} of the mode-locked structure in the I-V curves
is completely wiped out!  For $\nu=1/2$,
the solution reveals remaining structure, but the smeared current
plateaus
are {\it not} centered at integer multiples of $I= e \omega /2\pi$.

In Section V we compute the $I-V$ curves in a perturbative approach,
which leads us to conjecture the following general form
for the I-V curves at arbitrary $\nu$:
\begin{equation}
I(V_{sd}, V_{ac}) = \sum_n |c_n|^2 \; I^{dc}(\nu V_{sd} + n \omega). 
\end{equation}
Here $ I^{dc}(V) \equiv I(V,0)$ is the tunnel current
in the absence of the a.c. drive, and $|c_n|^2 = |J_n(\nu V_{ac}/ \omega)|^2$, with
$J_n(X)$
n'th order Bessel functions.  
These coefficients satisfy the sum rule
$\sum_n |c_n|^2 = 1$.  This form has a simple physical interpretation:
Charge $\nu$ quasiparticles absorb n-quanta from the a.c. field
with probability $|c_n|^2$, and are transmitted through the point contact
with total energy $\nu V_{sd} + n\omega$.

Equation (1.1), which is also consistent with our exact solutions, gives a simple
explanation as to why
all plateaus are wiped out at $\nu=1$.  For $\nu=1$
the edge states are describable in terms of non-interacting electrons
(Fermi liquid).
Under the assumption of an energy independent transmission
probability through the junction, the d.c. I-V
curves are linear (Ohmic).  Since the transmission is independent of energy,
the a.c. drive has {\it no} effect on the I-V curves, which remain 
completely linear.

For $\nu<1$ the d.c. I-V curves are non-linear, and plateau-like
features show up with an a.c. drive.  Recently, Fendley et. al.\cite{fendley}
have obtained exact solutions for the d.c. I-V curve
at arbitrary integer $\nu^{-1}$.  These curves, together with the conjecture
(1.1), enable us to construct the I-V curves with a.c. drive present
for the experimentally relevant cases of $\nu=1/3$ and $\nu=1/5$.
For these cases, in the limit of weak pinch off at the point contact,
the I-V curves exhibit smeared current plateaus centered at integer multiples
of $I=e \omega /2 \pi$.

Section VI is devoted to a discussion of the experimental
consequences.

\section{Model for Point contact with ac drive}

Consider then a FQHE state at filling $\nu^{-1}$
an odd integer.  For this class of Hall fluids only a single edge mode 
is expected\cite{wen}.  For the IQHE at $\nu=1$ a free-fermion description
of the edge mode is possible\cite{kane}, but more generally the edge mode is
expected to be a (chiral) Luttinger liquid, describable in terms
of a bosonic field.  

Let $\rho_R$ and $\rho_L$ denote the electron densities
in the right and left moving edge modes, on the top and bottom of the
sample, as shown schematically in Figure 1.  These densities are written
as gradients of bosonic fields,
\begin{equation}
\rho_{R/L} = {\pm}  \frac1{2\pi}  \partial_x \phi_{R/L}
\end{equation}
which satisfy the Kac-Moody commutation relations\cite{wen}:
\begin{equation}
[ \phi_{R/L} (x), \partial_x \phi_{R/L} (x') ]  = \mp i 2\pi \nu \delta(x-x')  .
\end{equation}
Here $x$ is a one-dimensional position coordinate, running along the edge.
The appropriate Hamiltonian density describing propagation of
edge modes is\cite{kane,moon}
\begin{equation}
{\cal H}_0 = \frac{v_F}{4 \pi\nu} [ (\partial_x \phi_R)^2 +(\partial_x \phi_L)^2]  .
\end{equation}
Here $v_F$ is the velocity of edge propagation.

\begin{figure}[hbt]
\epsfysize=2 truein
\epsfxsize=3 truein
\centerline{\epsfbox{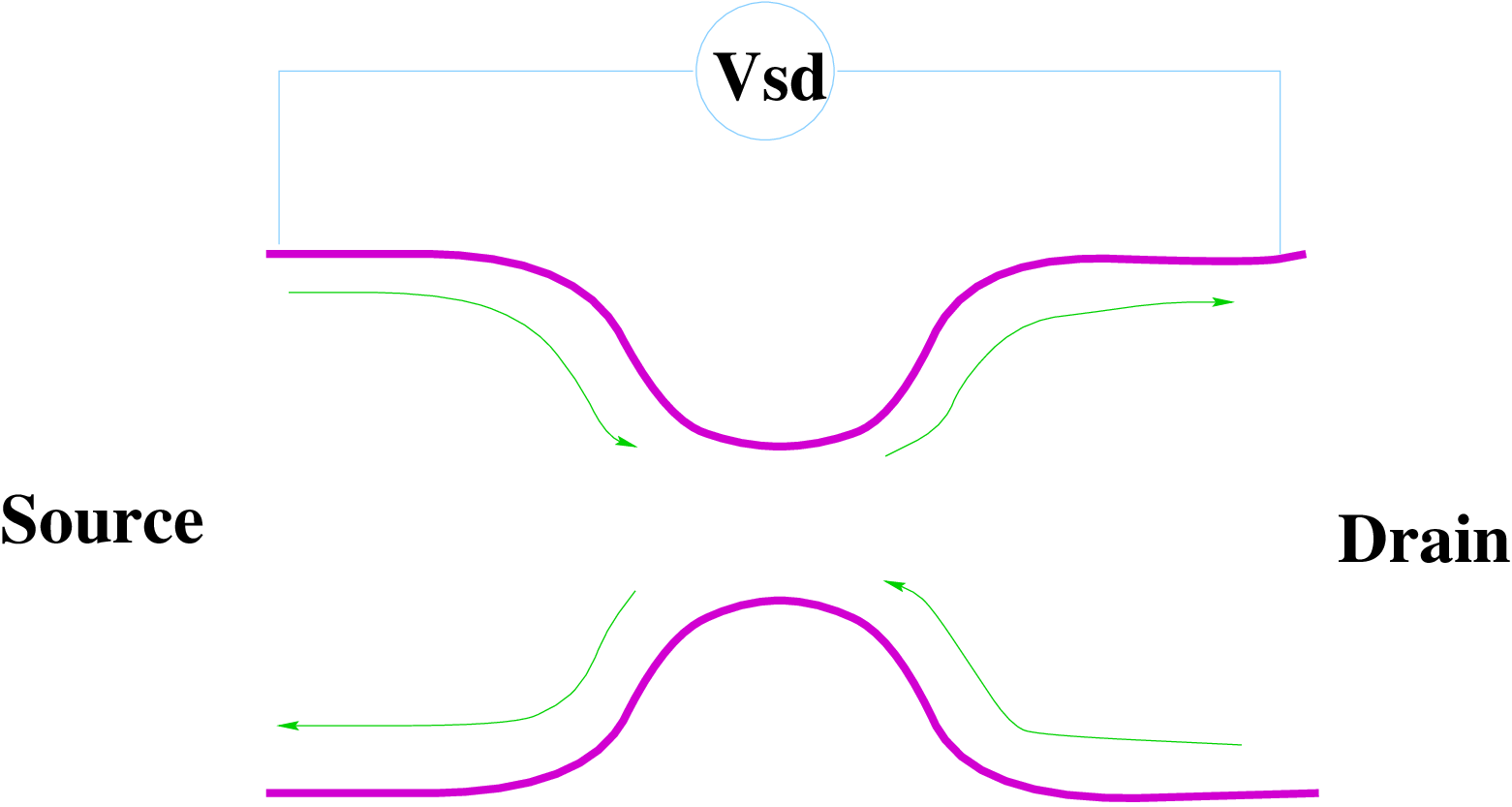}}
\vspace{.3in}
{\noindent FIG. 1:Schematic
representation of a point contact in a FQHE fluid.
The lines with arrows represent edge states which can scatter
at the point contact.  The voltage drop between source and drain
is denoted $V_{sd}$.
}
\end{figure}

At the point contact, the right and left moving edge modes
are brought into close proximity, and tunneling between them becomes
possible.  In the limit of weak tunneling,
the dominant backscattering process at low temperatures
is of fractionally charged ($Q=e\nu$) Laughlin quasiparticles\cite{stone}.
The appropriate tunneling term is
\begin{equation}
{\cal H}_1 = v \delta(x) \left[  e^{i(\phi_R - \phi_L)} + h.c. \right]  ,
\end{equation}
where $v$ is the local tunneling amplitude, at the point contact, $x=0$.

In the presence of an applied source-to-drain voltage,
the incoming edge modes will be at different chemical potentials.
Inter-edge tunneling processes will thus change
the energy.  Denoting the applied voltage as $V(t)$,
the energy change can be written: 
\begin{equation}
{\cal H}_V =  (\rho_R - \rho_L) \frac12 V(t) .
\end{equation}
In addition to a dc source-to-drain voltage, $V_{sd}$,
we will consider an applied ac field, arising from 
electro-magnetic radiation illuminating the point contact.
The total voltage drop between edges is written,
\begin{equation}
V(t) = V_{sd} + V_{ac} \sin \omega t  .
\end{equation}
For later convenience it will be useful to introduce a
gauge field A(t), defined via, $V(t)=\partial_t A(t)$.
A useful identity is 
\begin{equation}
e^{i\nu A(t)} = e^{i \nu V_{sd}t} \sum_{n=-\infty}^{\infty} c_n e^{-in\omega t}
, 
\end{equation} 
where $c_n = (-i)^n J_n(\frac{\nu V_{ac}}{\omega})$, with $J_n(X)$
Bessel functions.
The full Hamiltonian density is ${\cal H}={\cal H}_0+{\cal H}_1+{\cal H}_V$.

In the absence of backscattering at the point contact,
the total source-to-drain current, $I=v_F (\rho_R - \rho_L)$,
upon averaging over time 
is appropriately quantized: $<I>=\nu V_{sd}/2\pi$.  
Backscattering will reduce this current, 
\begin{equation}
<I>={1 \over {2\pi}} \nu V_{sd} - <I_B> ,
\end{equation}
where $I_B$ is the backscattering current operator.
An expression for $I_B$ follows upon functional differentiation:
\begin{equation}
I_B \equiv -\frac{\delta H}{ \delta A}
= \partial_t \frac12 \int dx (\rho_R - \rho_L).
\end{equation}

For later convenience it will be useful to define new boson fields
which propagate in the same
direction:
\begin{equation}
\phi_1(x) \equiv \phi_R(x), \hspace{.5in} 
\phi_2(x) \equiv \phi_L(-x)  .
\end{equation}
The commutators become,
\begin{equation}
[ \phi_i(x), \partial_x \phi_j(x')] = -i \delta_{ij} 2 \pi \nu \delta(x-x') .
\end{equation}
The Hamiltonian density has the same form as before,
\begin{eqnarray}
{\cal H} = \frac{v_F}{4\pi \nu}&& (\partial_x \phi_i)^2 
           + v \delta(x) \left[  e^{i(\phi_1 - \phi_2)} + h.c. \right]
\nonumber\\
        &&   +  (\rho_1 - \rho_2) \frac12 V(t) 
\end{eqnarray}
provided the densities are defined as,
$\rho_i = \partial_x \phi_i/2\pi$.
Upon using the  
continuity equations, $\partial_t \rho_i + v_F \partial_x \rho_i = 0$,
valid away from the point contact at $x=0$,
the backscattering current operator can be re-expressed as:
\begin{eqnarray}
I_B &&= - (v_F/2) \int dx\; \partial_x (\rho_1 - \rho_2)\nonumber\\
    &&= (v_F/2) (\rho_1 - \rho_2)\left| _{x=0^-}^{x=0^+}\right  .
\end{eqnarray}
Here we have used the fact that the only backscattering is at the origin,
$x=0$.

It is worth emphasizing that the above model
is only appropriate for a FQHE point contact
at filling factor $\nu^{-1}$
an odd integer.
For FQHE states at other filling factors, multiple
edge modes are expected.  Nevertheless, it will prove useful below to 
study the above model for arbitrary
$\nu$.

The current voltage characteristics of the point contact
follow upon computing the backscattering current
(2.13).
 Before attempting this, we consider briefly a semi-classical
limit of the model which
describes an overdamped driven classical pendulum.
Under exchange of current with voltage, this is identical
to the standard RSJ model of Josephson junction
dynamics\cite{mccumber,stewart,johnson}.
This classical model has been studied intensively, both because
of it's relevance to Josephson junctions, but also as a simple example
of a classical dynamical system which exhibits
mode locking and a devil's staircase\cite{jensen}.
  
\section{Semi-Classical limit}

To take the semi-classical limit we first
review the equivalence between the quantum Hall point contact
and the Caldeira-Leggett model\cite{caldeira} for the quantum mechanics
of a damped pendulum.  To this end, it is first useful to
perform a gauge transformation to eliminate ${\cal H}_V$ in (2.5).
Since the equations of motion for $x\ne 0$ take the form,
\begin{equation}
(\partial_t \pm v_F \partial_x)\phi_{R/L} = \pm \frac{\nu}{2} V(t) ,
\end{equation}
this can be achieved via the transformation:
\begin{equation}
\phi_{R/L} \rightarrow \phi_{R/L} \pm  \frac{\nu}{2}  A(t)  .
\end{equation}
After the gauge transformation, the full Hamiltonian reads,
\begin{eqnarray}
{\cal H} &=& \frac {v_F} {4\pi \nu}
\left[ (\partial_x \phi_R)^2 + (\partial_x \phi_L)^2 \right]
\nonumber\\
&+& v \delta(x)   \cos (\phi_R - \phi_L +\nu A)  .
\end{eqnarray}
Since the interaction term only depends on the difference,
$\phi_R - \phi_L$, it is useful to define new fields:
\begin{eqnarray}
\varphi &= \phi_R + \phi_L  ,\hspace{.3in}
\theta  &= \phi_R - \phi_L  .
\end{eqnarray}
Since the transformed Hamiltonian is quadratic in $\varphi$,
it can be integrated out, giving for
the Euclidean Lagrangian,
\begin{equation}
{\cal L}_E = \frac1{8 \pi \nu} (\partial_{\mu} \theta )^2 
+ v \delta(x) \cos (\theta + \nu A).
\end{equation}
Here we have set $v_F=1$ in the first term.
Finally, upon integrating out $\theta(x)$ for $x \ne 0$, 
we arrive
at an effective Euclidean action in terms of $\theta(x=0,\tau)$:
\begin{equation}
S_E = \frac1{4 \pi \nu} \int \frac{d\omega}{2 \pi}\; |\omega| |\theta(\omega)|^2
   +\int d\tau\; v \cos (\theta + \nu A ) 
\end{equation}
This action can be recognized as a Caldeira-Leggett model of
a damped driven quantum pendulum\cite{caldeira}.
It should be emphasized that the Ohmic damping that characterizes
the Caldeira-Leggett model can be traced to the 1d Luttinger liquid
behavior of
the edge modes.  Although this model
has been used to describe quantum dynamics in Josephson
junctions, it is unclear that it describes the appropriate low
frequency dynamics.  In particular, the phase of the 
Cooper pair field has long-ranged order
in the bulk superconducting electrodes,
in contrast to the power law correlations described by the 1d edge modes in (2.3).

Since we are interested in the non-equilibrium current-voltage
characteristics, we need a real time 
formulation, such as Keldysh\cite{fisher}.
In the Keldysh approach a generating functional is introduced
as a path integral sum over two paths
propagating forwards and backwards in time, $\theta_{\pm}(t)$:
\begin{equation}
Z = \int D\theta_+ D\theta_- \; e^{-S(\theta_{\pm})}  .
\end{equation}
In terms of new fields,
\begin{equation}
\theta(t) = \frac12 [ \theta_+(t) + \theta_-(t)], \hspace{.3in}
\tilde \theta (t)= \theta_+(t) - \theta_-(t)
\end{equation}
the appropriate real time action is $S = S_0 +S_1$ with,
\begin{equation}
S_0  = \frac12 \int d\omega \alpha_R(\omega) |{\tilde \theta}(\omega)|^2
        - \frac{i}{2 \pi \nu} \int dt\; {\dot \theta}{\tilde \theta}  ,
\end{equation}
\begin{equation}
S_1 = \sum_{\pm} \int dt 
(\pm iv) \cos ( \theta \mp \frac12 {\tilde \theta} + \nu A )  .
\end{equation}
Here we have defined $\alpha_R(\omega) = 
\frac{\omega}{2\pi \nu} \coth (\frac12 \beta \omega)$.
The above gives a general quantum-mechanical formulation
of the model.  To complete the description we must identify the source-to-drain
current operator.  From (3.4) we see that
$\theta(x=0)=2 \pi \int_{- \infty}^0 \rho_{tot} dx$, where
$\rho_{tot} = \rho_R + \rho_L$.  Thus $\theta(x=0)/2\pi$ can be identified
as the total charge to the left of the point contact.
The source-to-drain current through the point contact is thus simply:
\begin{equation}
I = \partial_t \theta(x=0,t)/2\pi  .
\end{equation}

An instanton in $\theta(t)$ of magnitude
$2 \pi$ corresponds to the transfer of one electron through
the point contact.  In the classical limit these 
charge transfer processes occur over the barrier,
rather than by quantum mechanical tunneling.
In the Keldysh formulation, quantum tunneling processes
correspond to instantons in $\tilde {\theta(t)}$  - in which
only the forward path tunnels, say.  Thus the semi-classical
limit can be obtained by forbidding such processes.
This can be implemented by expanding the cosines in (3.10)
for small $\tilde {\theta}$, and retaining only the leading term
\begin{equation}
S_1  = iv \int dt\; \tilde \theta\: \sin(\theta + \nu A) 
    + O(\tilde \theta^3)  .
\end{equation}
This expansion destroys the periodicity in $\tilde {\theta}$.
The full action can now be written
\begin{eqnarray}
S =&& \frac12 \int d\omega\; \alpha_R(\omega) |{\tilde \theta}(\omega)|^2
\nonumber\\
&&-i \int dt\; \tilde \theta \left[ \frac1{2 \pi \nu} \dot \theta
 - v\sin(\theta + \nu A) \right]  ,
\end{eqnarray}
which can be recognized as the Martin-Siggia-Rose action
for a classical stochastic differential equation\cite{martin}.
Upon introducing a stochastic noise term
$\xi(t)$ the action can be re-expressed as
\begin{eqnarray}
S =&& \frac12 \int d\omega \;\frac{1}{\alpha_R(\omega)}|\xi(\omega)|^2
\nonumber\\
&&-i \int dt\; \tilde \theta \left[ \frac1{2\pi \nu} \dot \theta
 - v\sin(\theta + \nu A)+ \xi(t) \right] .
\end{eqnarray}
The integration over $\tilde {\theta}$ then gives a delta function,
enforcing the classical equation of motion:
\begin{equation}
\frac1{2\pi \nu}\dot \theta=  
 v\sin(\theta + \nu A)+ \xi(t)  ,
\end{equation}
with stochastic noise
\begin{equation}
\left< |\xi(\omega)|^2 \right> = 
\frac{\omega}{2\pi \nu} \coth (\frac12 \beta \omega) .
\end{equation}
A final gauge transformation, $\theta \rightarrow \theta - \nu A$
brings the equation into the familiar form,
\begin{equation}
\frac1{2\pi}\dot \theta =  
 \nu v\sin(\theta) + \frac{\nu}{2\pi} V(t) + \nu \xi(t).
\end{equation}

\begin{figure}
\epsfysize=3.5 truein
\epsfxsize=3 truein
\centerline{\epsfbox{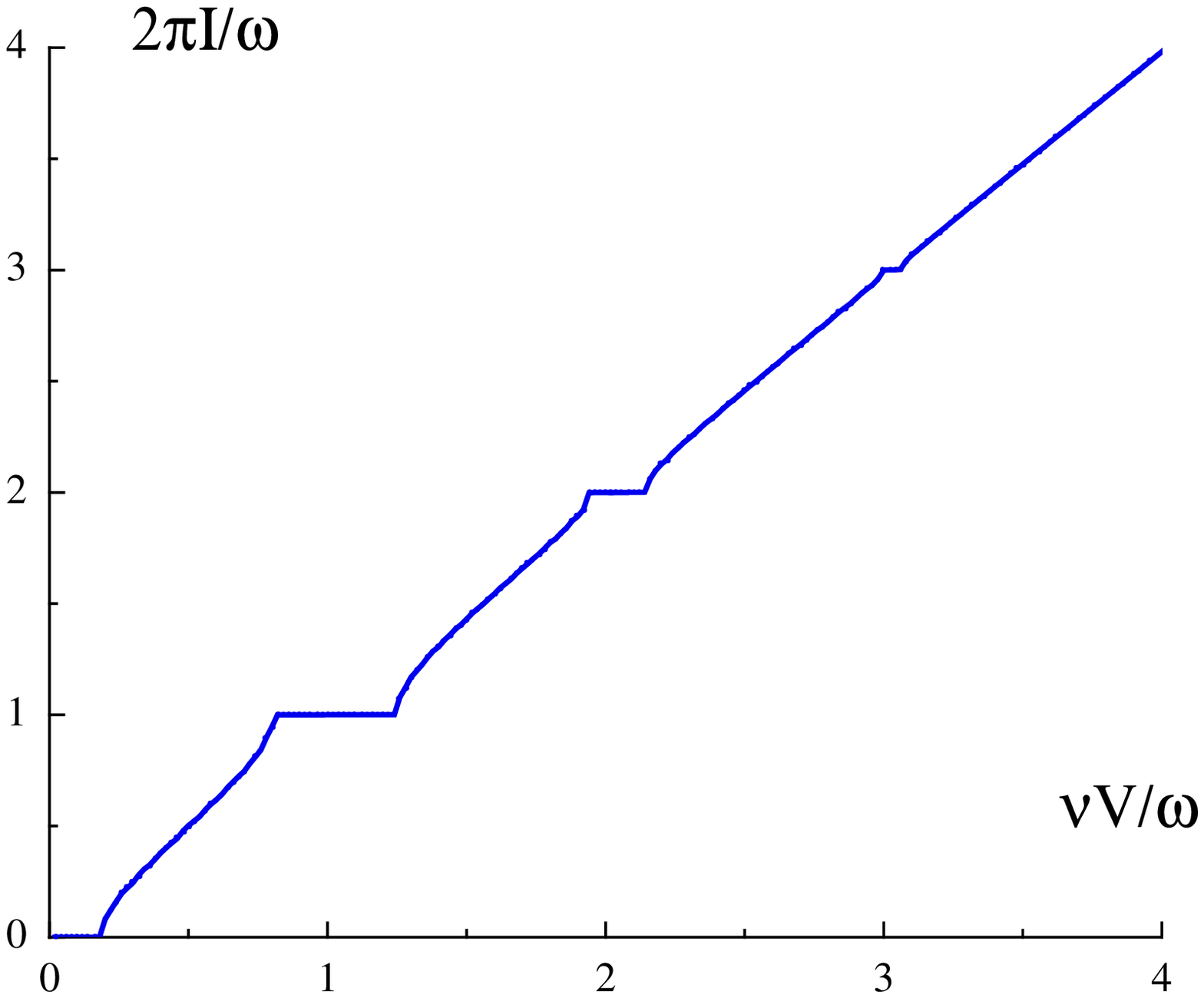}}
{\noindent FIG. 2:
Current voltage characteristic within the semi-classical approximation,
obtained from (3.17) with no stochastic noise. Both the current and voltage
are plotted in units of the ac drive frequency.  
The backscattering amplitude has been chosen
to be $2\pi \nu v = \omega/4$ and the ac drive amplitude is 
$\nu V_{ac}=1.6\omega$.
Notice the current plateaus at integer multiples
of $ I = e\omega/2\pi$, indicating a mode-locking to the ac drive. 
}
\end{figure}

Under exchange of current and voltage, (3.17) becomes
equivalent to the equation which describes Josephson
junctions\cite{mccumber,stewart,johnson}, except for
the colored stochastic noise term which is non-vanishing even at
zero temperature.  However, if we take the semi-classical limit
$\nu \rightarrow 0$, with $\nu v$ and $\nu V(t)$ held fixed,
the noise term drops out.
In this classical limit, the FQHE point contact is exactly dual to
a Josephson junction, and should exhibit similar mode-locking under
exchange of current and voltage.
Solutions of (3.17) in the noiseless limit
are well known\cite{hamilton,russer,hahlbohm}. 
For a Josephson junction they give mode-locked voltage
plateaus at integer multiples of $V=(\hbar /2e)\omega$.
Physically, there is a mode locking between the discrete phase slip events
and the a.c. drive.
For the FQHE point contact,
the mode-locked plateaus are in the current, at integer multiples
of $I=e2\pi \omega$.  The discrete process is an electron
tunnelling through the point contact.

After rescaling the time
in (3.17) via $t \rightarrow \omega t$,
it is clear that the I-V curves are characterized
by two independent dimensionless
parameter:
$2\pi \nu v/\omega$ and
$\nu V_{ac}/\omega$. 
Representative current-voltage characteristics computed numerically
from (3.17) in the noiseless limit are shown in Figure 2.
As expected, the $I-V$ curves exhibit plateaus in the current
which are ``flat" and quantized at integer multiples
of $I=e\omega/2\pi$.  
Sub-harmonic plateaus are absent for the model (3.17),
but would be present if the periodic function $sin(\theta)$
included higher harmonic content\cite{azbel}.

With inclusion of stochastic noise,
one anticipates that these plateaus will be rounded slightly,
as shown in the I-V curves in Figure 3, obtained
by numerically 
integrating (3.17) with colored noise.
When the noise is weak, the rounding
is most visible at the edges of the plateaus.
For large enough noise the plateaus become completely smeared out.
The effects of colored noise are qualitatively similar to stochastic white
noise, which has been studied extensively in the past.

It is worth commenting here on the validity of the semi-classical
approximation to the full quantum dynamics.
As evident from (3.12), the semi-classical approximation involves
discarding all electron tunnelling events, in which $\tilde{\theta}$
changes by $2 \pi$.  One can argue from the quadratic action (3.9)
that the typical variance of $\tilde{\theta}$ is proportional to $\nu$,
even when $v=0$: $\tilde{\theta}^2 \sim \nu ln(\omega_c/T)$,
with cutoff frequency $\omega_c$.  This suggests that the semi-classical
expansion in (3.12)
might become exact in the $\nu \rightarrow 0$ limit. 
In the absence of an ac drive this is in fact the case. 
Recently, Fendley et. al. have obtained exact $I-V$ curves,
with no ac drive, for arbitrary odd integer $\nu^{-1}$. 
One can analyze these $I-V$ curves in the limit $\nu \rightarrow 0$,  
with $\nu v$ and $\nu V$ 
held fixed.  In this limit, the $I-V$ curves become equivalent to
those which follow from the classical equation of motion (3.17),
with white noise replacing the stochastic colored noise.

\begin{figure}
\epsfysize=3.5 truein
\epsfxsize=3 truein
\centerline{\epsfbox{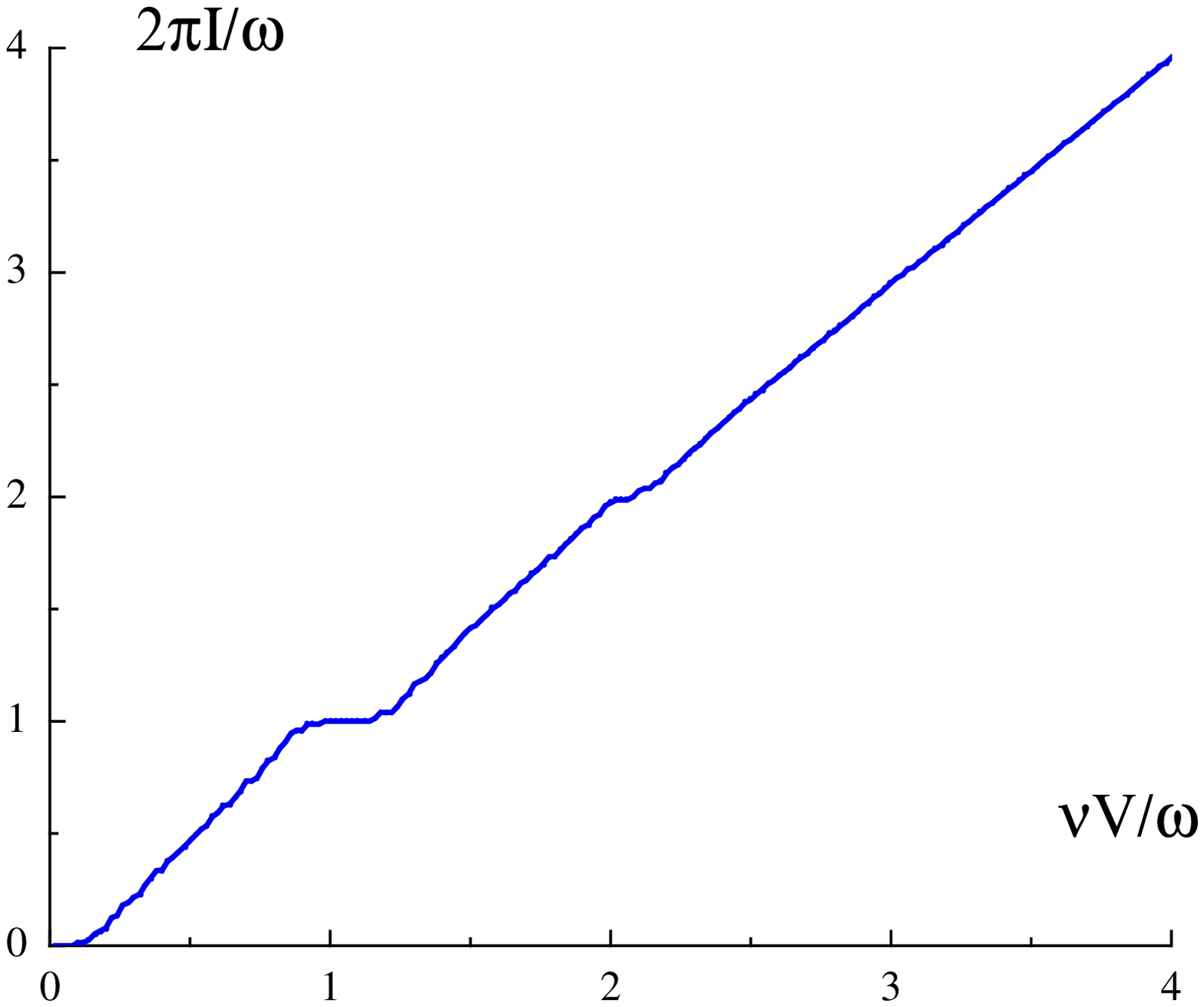}}
{\noindent FIG. 3:
An $I-V$ curve in the semi-classical approximation, obtained
from (3.17) with stochastic colored noise.  As in Figure 2,
we choose $2 \pi \nu v = \omega/4$ and $\nu V_{ac}=1.6\omega$. 
 The colored noise has strength $\nu=0.1$ and the cut-off frequency for 
generating the noise is $\omega_c = 60 \omega$.
Notice that the current plateaus are rounded, due to the presence
of the stochastic noise.
}
\end{figure}

However, with an ac drive present, it is unlikely that the $\nu \rightarrow
0$ limit is equivalent to the semi-classical limit (3.17).
With ac drive present there are two parallel processes which allow charge to
be transported across the junction.  In addition to electron tunnelling
``under the barrier", the electron can absorb quanta of energy from the ac drive
field.  Once the electron energy is high enough, it can pass
over the washboard barrier.  In the classical limit, both of these processes
are modified:  Electron tunnelling is suppressed completely, and
energy is not absorbed from the ac drive in discrete quanta.
However, in the $\nu \rightarrow 0$ limit,
while the electron tunnelling is also completely suppressed
(since  $I \sim V^{2/\nu -1} \rightarrow 0$ as $\nu \rightarrow 0$),
energy is still absorbed in discrete quanta from the ac drive.
Thus, once the ac drive is present, one anticipates that
mode locking features obtained from the semi-classical limit (3.17),
will {\it not} serve as a good guide for the full quantum model,
even for very small $\nu$.  This will be confirmed by more detailed 
analysis in Section V below.

\section{ Exact solutions for $\nu=1, \frac12$}

In this section we study
the full quantum dynamics for two special values of $\nu$,
for which simple exact solutions are possible.
For $\nu=1$ the edge mode is equivalent to a free fermion\cite{kane}.
When the a.c. drive is present, an exact solution for the
I-V curve is possible.
When $\nu=1/2$ a free fermion representation is also possible\cite{fendley1}.
Although the theory is not directly applicable to
the FQHE edge states for $\nu=1/2$, the exact solution is nevertheless
illuminating, revealing plateau-like structure
in the I-V curve, in contrast to $\nu=1$
(see below).  Moreover, the general structure of the solutions
in these two soluble cases, leads to a natural conjecture
for more general $\nu$, discussed in Section VI.

\subsection{The $\nu=1$ solution}
For $\nu=1$ the edge modes have a free fermion description,
simpler than the general bosonized representation of Section II.
Upon defining fermion fields for the two modes,
\begin{equation}
\Psi = \left( \begin{array}{c}
              \psi_1\\  \psi_2
              \end{array}  \right)
     = \frac{1}{\sqrt{a_0}}\left( \begin{array}{c}
              e^{i\phi_1}\\  e^{i\phi_2}
              \end{array}  \right)
\end{equation}
with 
$a_0$ a short length-scale cutoff, the full bosonized Hamiltonian
can be expressed as a quadratic fermion theory:
\begin{equation}
{\cal H} = -\Psi^{\dag} \left[ i\partial_x + \frac12 V(t) \mbox{ \boldmath
$\sigma$}_z \right]
           \Psi + \frac{v}{\omega_c} \delta(x) \Psi^{\dag} \mbox{ \boldmath
$\sigma$}_x \Psi  .
\end{equation}
Here, we have put the Fermi velocity $v_F =1$, and the cutoff frequency 
$\omega_c \sim
\frac{1}{a_0}$.
The backscattered current (2.13) takes the simple form:
\begin{equation}
I_B = (1/2) \Psi^{\dag} \mbox{ \boldmath $\sigma$}_z \Psi \left|
_{x=0^-}^{x=0^+}\right .
\end{equation}

The equation of motion which follows from the fermion Hamiltonian is
\begin{equation}
\left[ \partial_t + \partial_x - \frac{i}2 V(t) \mbox{ \boldmath $\sigma$}_z
\right] \Psi
= -i \frac{v}{\omega_c} \delta(x) \mbox{ \boldmath $\sigma$}_x \Psi  .
\end{equation}
The I-V curve can be obtained by solving this equation, with
appropriate boundary conditions, and extracting the backscattered
current, $I_B$.

Our solution proceeds in two steps.  Away from the point contact
at $x=0$, the equation describes free propagation with
a uniform time-dependent potential, $V(t)$.  This can be eliminated
by defining a new gauge transformed fermion field,
which is assumed to be incident upon the point contact
with a Fermi-Dirac distribution.  Upon transforming back to the original
Fermion field, the Fermi distribution function is modified,
involving a sum over processes involving absorption and emission of
the a.c. field.  
We refer to this distribution as an ``excited Fermi function".
At the point contact ($x=0$), backscattering takes place,
which is characterized by reflection and transmission coefficients (an S-matrix)
which are independent of the incident distribution function.
The total backscattered current, $I_B$, is an appropriate convolution
of the $S-matrix$ with the ``excited Fermi" distribution function. 

Consider first scattering at the point contact.
The $\bf S$ matrix relates the incoming
field $\Psi_-$ to the  
outgoing field $\Psi_+$ via
\begin{equation}
\Psi_+ = {\bf S} \psi_-  ,
\end{equation}
where $\Psi_{\pm}(t) = \Psi(x=0^{\pm},t)$.
Integrating the equation of motion (4.4) through the origin,
$x=0$, gives,
\begin{equation}
\Psi_+(t)  -  \Psi_-(t) = -i \frac{v}{\omega_c}
 \mbox{ \boldmath $\sigma$}_x \Psi(0,t),
\end{equation}
where $\Psi(0,t) = \frac12[\Psi_+(t) + \Psi_-(t)]$.\\
From this one readily obtains the $\bf S$ matrix:
\begin{equation}
S_{11} = S_{22} = 
\frac{ 1-(v/2 \omega_c)^2}{1+(v/2 \omega_c)^2}, 
\end{equation}
\begin{equation}
S_{12}=S_{21} = \frac{-iv/ \omega_c}{1+(v/2 \omega_c)^2}.
\end{equation}
The probability for the incoming field
to be scattered from one edge to the other
is $|S_{12}|^2$, whereas $|S_{11}|^2$ is the probability
to be transmitted without scattering.
Probability conservation dictates a unitary S-matrix,
${\bf S^\dagger} {\bf S} = {\bf 1}$, which is satisfied here.
Notice that the S-matrix is independent of the
energy of the incident carriers, a consequence of the assumed
delta-function point scatterer.

Outside the scattering region, the right side of (4.4) vanishes.
Transforming to a new fermion field
\begin{equation}
\Psi(x,t) = e^{\frac{i}2 A(t) \mbox{ \boldmath $\sigma$}_z} 
\tilde{\Psi}(x,t)  ,
\end{equation}
with $V(t) = \partial_t A(t)$ as before, then eliminates the time dependence.
The new field satisfies the simple wave equation: $(\partial_t + \partial_x)
\tilde{\Psi} = 0$, which describes free fermions at
zero chemical potential.  This field is assumed
to be incident with an ordinary Fermi distribution function,
\begin{equation}
 \left< \tilde{\psi}_i(E)^{\dag} \tilde{\psi}_i(E') \right>  = 
2 \pi \delta(E-E') f(E) , 
\end{equation}
where $f(E) = (exp(\beta E) + 1)^{-1}$ and
$\tilde{\Psi}(E)$ denotes the Fourier transform
of $\tilde{\Psi}(x=0^{-},t)$.

The distribution function for the original incident Fermion,
$\Psi_-(t)$, can now be obtained by relating the
 transform $\Psi_-(E)$ to $\tilde {\Psi}(E)$ using (4.9) 
and the expansion (2.7).  This gives
\begin{equation}
\psi^-_{1,2}(E) = \sum_n c_n \tilde{\psi}_{1,2} (E-n \omega \pm \frac12 V_{sd}),
\end{equation}
where $c_n$ is defined in (2.7).
The distribution function for the original Fermion,
$ < \psi^{-\dagger}_{j}(E) \psi^-_{j}(E')> = 2\pi \delta(E-E') f^{ex}_{j}(E)$,
then takes the simple
form:
\begin{eqnarray}
f^{ex}_{1,2} = \sum_n |c_n|^2 f(E-n \omega \pm \frac12 V_{sd}),
\end{eqnarray}
an ``excited Fermi function".  
Notice that the d.c. voltage $V_{sd}$ simply causes
a shift in the energy of the incident electron.
The a.c. drive shifts the energy by $n \omega$, 
corresponding to absorption or emission of $n$ quanta,
with probability $|c_n|^2$.

Finally we can obtain the backscattered current from (4.3), which can be
re-expressed using (4.5) solely in terms of the incident fields as
$I_B  = - |S_{12}|^2  
\Psi_-^{\dag} \mbox{ \boldmath $\sigma$}_z \Psi_-$.
After Fourier transforming to
energy this becomes,
\begin{equation}
\left<I_B \right>= -\int_{E,E'} e^{-i(E-E')t}  |S_{12}|^2 
\left< \Psi_-^{\dag}(E') \mbox{ \boldmath $\sigma$}_z \Psi_-(E) \right>.
\end{equation}
In addition to a time-independent piece, the backscattered current
will have oscillatory contributions at multiple frequencies
of $\omega$, as is apparent from (4.13).  We focus only on the time independent
piece, which is finally given by
\begin{equation}
\left< I_B \right>_{time}= \frac{1}{2 \pi}
\int dE \: |S_{12}|^2 [ f^{ex}_2(E) - f^{ex}_1(E) ].
\end{equation}
This result takes a familiar form, involving an energy
integral of the reflection probability, weighted
by
energy distribution functions.  Due to the a.c. drive, however,
these are not simply Fermi functions, but rather
the ``excited Fermi functions" given in (4.12).

Since the reflection probability is energy independent,
the backscattered current can be seen to be completely
independent of the a.c. drive.  This follows
by inserting the distribution function, (4.12), 
and shifting the energy of integration to eliminate
the drive frequency $\omega$.  Since $\sum_n |c_n|^2=1$,
the backscattered current is then exactly equal to the result
without any a.c. drive present.  At zero temperature 
this gives $<I_B>_{time} = (1/2\pi)|S_{12}|^2 V_{sd}$,
or for the total transmitted current (using (2.8)) upon
restoring units:
\begin{equation}
 I = \frac{e^2}{h} |S_{11}|^2 V_{sd}  .
\end{equation}
The I-V curve is linear, with conductance given by the transmission
probability, just as without any a.c. drive.
The quantum fluctuations have completely washed out
the current plateaus seen in the semi-classical
limit of Section III.  The absence
of structure in the I-V curve can be traced to the energy independent
transmission probability.  The a.c. drive
changes the energy of the incident electron, via
absorption or emission of quanta, but since
the transmission
probability is energy independent, this has no effect
on the net transmitted current.

It is worth mentioning that the total transmitted current
can be cast into the suggestive form:
\begin{equation}
I( V_{sd}, V_{ac}) = \sum_n |c_n|^2\; I^{dc}(\nu V_{sd}-n \omega),
\end{equation}
where $ I^{dc}(V) \equiv I(V,0)$ is the current in absence of a.c. drive.
As we shall now show, this form also holds when $\nu=1/2$,
even though in that case $I^{dc}(V)$ shows non-Ohmic structure.
Moreover, as discussed in Section V, this form is also
valid perturbatively in the weak backscattering limit,
for general $\nu$. 

\subsection{the $\nu= \frac12$ solution}

Consider now the model (2.12) with $\nu=1/2$.
In this case one can show using the commutation relations
(2.11), that the operator $exp(\phi_1 - \phi_2)$, that enters
in the Hamiltonian, satisfies Fermi statistics.
In order to fermionize this operator, it is convenient to define
new boson fields\cite{fendley,fendley1}:
\begin{eqnarray}
\phi(x,t) && =  [ \phi_1(x,t) - \phi_2(x,t)],\\
\Phi(x,t)&& =  [ \phi_1(x,t) + \phi_2(x,t)].
\end{eqnarray}
When the Hamiltonian (2.12) is re-expressed in terms of these
new fields, the field $\Phi$ decouples and can be ignored.
The remaining Hamiltonian becomes,
\begin{eqnarray}
{\cal H} = \frac {v_F}{4\pi} &&(\partial_x \phi)^2 
           + v \delta(x) \left[  e^{i\phi} + h.c. \right]\nonumber\\
          && +  \frac12 V(t) ( \partial_x \phi),
\end{eqnarray}
where we have set $\nu=1/2$.  

Since $e^{i\phi}$ has Fermi statistics, we can fermionize the remaining
boson field, via 
$ \Psi = \frac{1}{\sqrt{a_0}}
e^{i\phi}$, with lattice cutoff $a_0$.
The first term describes a free chiral fermion and the third term is also
quadratic in $\Psi$, however the tunnelling term is {\it linear}
in $\Psi$.  To convert this term into a quadratic form,
we introduce a local fermion field $a$ as,
\begin{equation}
\Psi(x) = (a +a^{\dag}) \psi(x) ,
\end{equation}
where both $a$ and $\psi(x)$ satisfy fermion anti-commutation relations.
The full Hamiltonian then becomes,
\begin{equation}
{\cal H}= \psi^{\dag} (i\partial_x + \frac12  V(t) ) \psi
 + \frac{v}{\sqrt{\omega_c}}\delta(x)
   [\psi^{\dag}(a+a^{\dag}) + h.c.].
\end{equation}
Here we have set $v_F=1$, and the cutoff frequency $\omega_c \sim 1/a_0$.
To complete the fermionization, we re-express the backscattering 
current from (2.13) in terms of the fermion fields: 
\begin{equation}
I_B = \frac12 \psi^{\dag}\psi \left|_{x=0-}^{x=0+}\right.
\end{equation}
Since the Hamiltonian (4.21) is quadratic, it can be readily solved,
and the current computed, as we now show.

To this end, consider first the equations of motion for the fermion fields
which follow from the Hamiltonian.  The local fermion satisfies
\begin{equation}
\partial_t (a + a^{\dag}) = 2 i \frac{v}{\sqrt{\omega_c}}
[ \psi(0) - \psi^{\dag}(0)],
\end {equation}
with $\psi(0) = (\psi(x=0^+)+\psi(x=0^-))/2$, whereas $\psi(x,t)$ satisfies,
\begin{equation}
\left[\partial_t + \partial_x - \frac{i}2 V(t)\right] \psi 
= i \frac{v}{\sqrt{\omega_c}} \delta(x)(a + a^{\dag}).
\end{equation}

We now proceed by direct analogy with the $\nu=1$ case.
Away from the point contact, the right side of (4.24)
vanishes, and the time dependent potential $V(t)$ can be eliminated
by gauge transforming to a new field.  At the point contact,
we compute the S-matrix, which relates
the amplitude of the incoming fermion 
($x=0^-$) to the outgoing
fermion ($x=0^+$).

To compute the S-matrix,
first integrate (4.24) through the origin ($x=0$),
and then eliminate the local fermion term $a+a^\dagger$
using (4.23).  This gives the local equation, 
\begin{eqnarray}
\partial_t (\psi_+ - \psi_-) &&= \frac{v^2}{\omega_c}
\; [ \psi_+^{\dag} + \psi_-^{\dag}-\psi_+ -  \psi_-],
\end{eqnarray}
where we have defined incoming and 
outgoing fields $\psi_\pm (t) = \psi(x=0^\pm ,t)$.
This can be converted to an algebraic equation by Fourier
transformation:
\begin{eqnarray}
\psi_+(E) - \psi_-(E) = &&\frac{v^2}{iE \omega_c}
   [\psi_+(E)+ \psi_-(E) \nonumber\\
&&-\psi_+^{\dag}(-E) - \psi_-^{\dag}(-E)].
\end{eqnarray}
Upon combining this equation with it's Hermitian conjugate,
we can eliminate
$\psi_+^{\dag}(-E)$, and express the outgoing field $\psi_+(E)$
in terms of the incoming fields $\psi_-(E)$ and $\psi_-^{\dag}(-E)$,
\begin{equation}
\psi_+(E) =  S_{++}(E) \;\psi_-(E) + S_{+-}(E)\; \psi_-^{\dag}(-E).
\end{equation}
Here the
energy dependent S-matrix elements are given by
\begin{equation}
S_{++}(E) = \frac{\alpha_E}{\alpha_E +i}\; , \hspace{.2in}
S_{+-}(E) = \frac{i}{\alpha_E + i},
\end{equation}
with  $\; \alpha_E \equiv \frac {E \omega_c}{2v^2}$.
As required by current conservation, the S-matrix satisfies
$\;\;| S_{++}(k)|^2 + |S_{+-}(k)|^2 =1\;\;$.

To obtain the distribution function for the incident fermion, we follow the
procedure used for $\nu=1$, and define a new fermion field
which eliminates the time dependent potential in (4.24):
\begin{equation}
\psi(x,t) = e^{\frac{i}2 A(t)}\; \tilde{\psi}(x,t).
\end{equation}
with $V=\partial_t A$.  After Fourier transformation this becomes,
\begin{equation}
\psi_-(E) = \sum_n c_n \tilde{\psi}_-(E - n \omega + \frac{V_{sd}}{2}). 
\end{equation}
Since 
the new field, $\tilde{\psi}$, satisfies the free wave equation, $(\partial_t +
\partial_x)\tilde{\psi}=0$,
for $x<0$, we assume again that it is incident upon the point contact
with a Fermi distribution function, $f(E)=(exp(\beta E)+1)^{-1}$.
The distribution function for the original fermion,
$<\psi_-^{\dag}(E) \psi_-(E')> =2 \pi \delta(E-E') f^{ex}(E)$,
is thus given again by the ``excited fermi function":
\begin{equation}
f^{ex}(E) = \sum_n |c_n|^2 
f(k-n\omega + \frac{V_{sd}}{2}).
\end{equation}

Finally, the backscattered current averaged over time follows from (4.22)
as,
\begin{equation}
\left< I_B \right>_{time} = \frac12 \int \frac{dE}{2 \pi}\;
\left< \psi_+^{\dag}(E) \psi_+(E) -  \psi_-^{\dag}(E) \psi_-(E) \right>.
\end{equation}
After re-expressing the outgoing waves in terms of incoming,
using the S-matrix (4.28), the averages over the incident
distribution can be performed, giving
\begin{equation}
<I_B>_{time} =  \int \frac{dE}{2 \pi}\; |S_{+-}(E)|^2
(\frac12- f^{ex}(E) ).
\end{equation}
The total transmitted current (2.8) can once again be cast into the
form:
\begin{equation}
I(V_{sd},V_{ac}) = \sum_n |c_n|^2 I^{dc}(\nu V_{sd}+n\omega)  ,
\end{equation}
with $\nu =1/2$.
Here the current in the absence of a.c. drive,
$I^{dc}(V_{sd}) \equiv I(V_{sd},0)$, is given by
\begin{eqnarray}
I^{dc} (V_{sd}) &=& {1 \over {4\pi}} V_{sd} 
- \int {{dE}\over{2\pi}} \bigg\{ |S_{+-}(E)|^2
\nonumber\\
&&(\frac12 - f(E+\frac12 V_{sd}))   .
\end{eqnarray}

\begin{figure}
\epsfysize=3.5 truein
\epsfxsize=3 truein
\centerline{\epsfbox{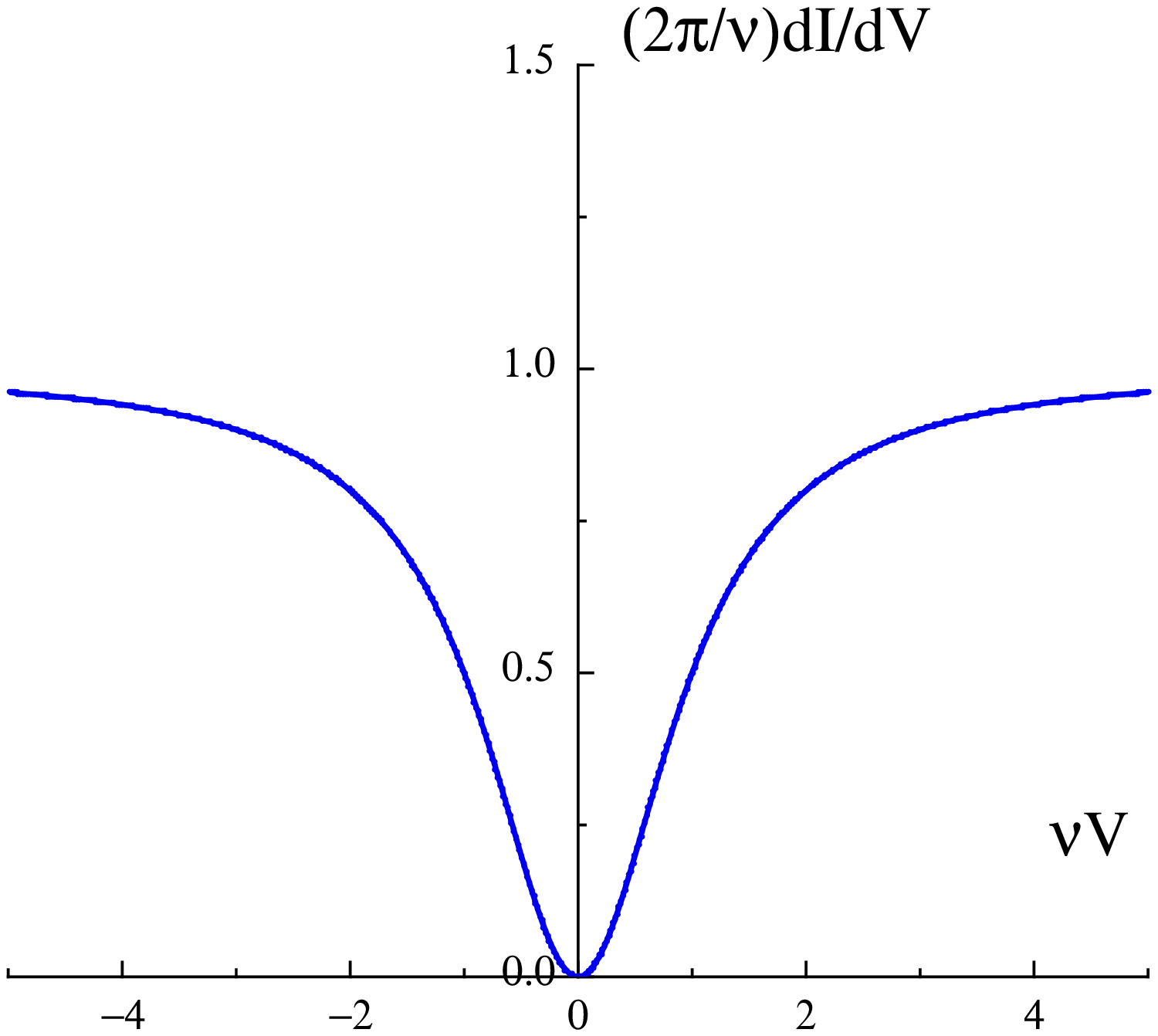}}
{\noindent FIG. 4:
Differential conductance with no ac drive at $\nu=1/2$.
We choose $\nu T_B = 1$ in this plot. 
}
\end{figure}

Notice that in contrast to the case $\nu=1$, the S-matrix here
depends on the energy of the incident fermion.  As a result,
the I-V curve is non-Ohmic.  The differential conductance 
at zero temperature as obtained from (4.35) in the absence of the
a.c. drive is plotted in Figure 4.  At small 
bias, the current vanishes with the cube of the source-drain voltage.
This is consistent with the general result,
$I \sim V_{sd}^{2/\nu -1}$, obtained from perturbation theory
in the strong backscattering limit.
For large bias the I-V curve is linear with an offset voltage.
Again, this is consistent with the general
perturbative result for small backscattering;
$((\nu/2\pi)V_{sd} - I) \sim v^2 V_{sd}^{2\nu -1}$.
Notice that for all $\nu <1/2$, the I-V curve at large
voltage is thus expected to asymptote to $I=(\nu/2\pi) V_{sd}$, with
{\it no} offset.  

With the a.c. drive present, the I-V curve can be obtained
by summing in (4.34), with weighting $|c_n|^2\:=\:J_n^2(
\frac{V_{ac}}{2\omega})$. 
Since the I-V curve with no a.c. drive is non-Ohmic at small bias,
this will give features in the full I-V curve which resemble
smeared current plateaus.
These plateaus can be more readily revealed by plotting
$dI/dV$ versus $V_{sd}$.

Anticipating the analysis of the I-V curves for general $\nu$
in the next section, it is convenient at this stage to define
an effective backscattering energy or temperature
scale.  Following Fendley et. al.,
we define a backscattering temperature
$T_B = g(\nu)\omega_c (v/\omega_c)^{1/(1-\nu)}$, where the function $g(x)$ is,
\begin{equation}
g(x) = \frac{4\sqrt{\pi}}{x} x^{\frac1{2-2x}}(\frac1x-1)^{\frac12}
       \frac{\Gamma(\frac1{2-2x})}{\Gamma(\frac{x}{2-2x})}  .
\end{equation}
In the $\nu \rightarrow 0$ limit,
one has $T_B = 2\pi v$, which is the appropriate backscattering
energy scale entering in the semi-classical equations of motion
(3.17). For $\nu=1/2$, $T_B = 4 v^2/\omega_c$ which is the energy
scale that enters in (4.28).
The $I-V$ curves 
at $T=0$ are then characterized by two
dimensionless parameters, $\nu \tilde{V}_{ac} = \nu V_{ac}/\omega$ and
$\nu \tilde{T_B} = \nu T_B/\omega$.

In Figure 5 we plot the differential conductance versus
voltage, obtained from (4.34)
with 
$\nu \tilde{V}_{ac} = 1.6$ and $
\nu \tilde{T_B} = 1/4$.
Notice the minima, which correspond to smeared plateaus in the $I-V$
curves.  The differential conductance at the nth minima is 
$ 1-J^2_n(V_{ac}/2\omega)$.  The widths of the minima
depend primarily on the backscattering energy scale,
$T_B$, becoming narrower as $T_B$ decreases.
In contrast to the semi-classical 
approximation (3.17), the ``plateaus" here have 
a non-vanishing differential conductance everywhere - they are not ``flat". 
Moreover, the I-V curve is everywhere analytic,
even at the ``plateau" centers, since
the 
I-V curve without a.c. drive is analytic
even at zero bias.  Evidently, quantum fluctuations are
quite effective at smearing the semi-classical current plateaus,
even for $\nu=1/2$.

\begin{figure}
\epsfysize=3.2 truein
\epsfxsize=3 truein
\centerline{\epsfbox{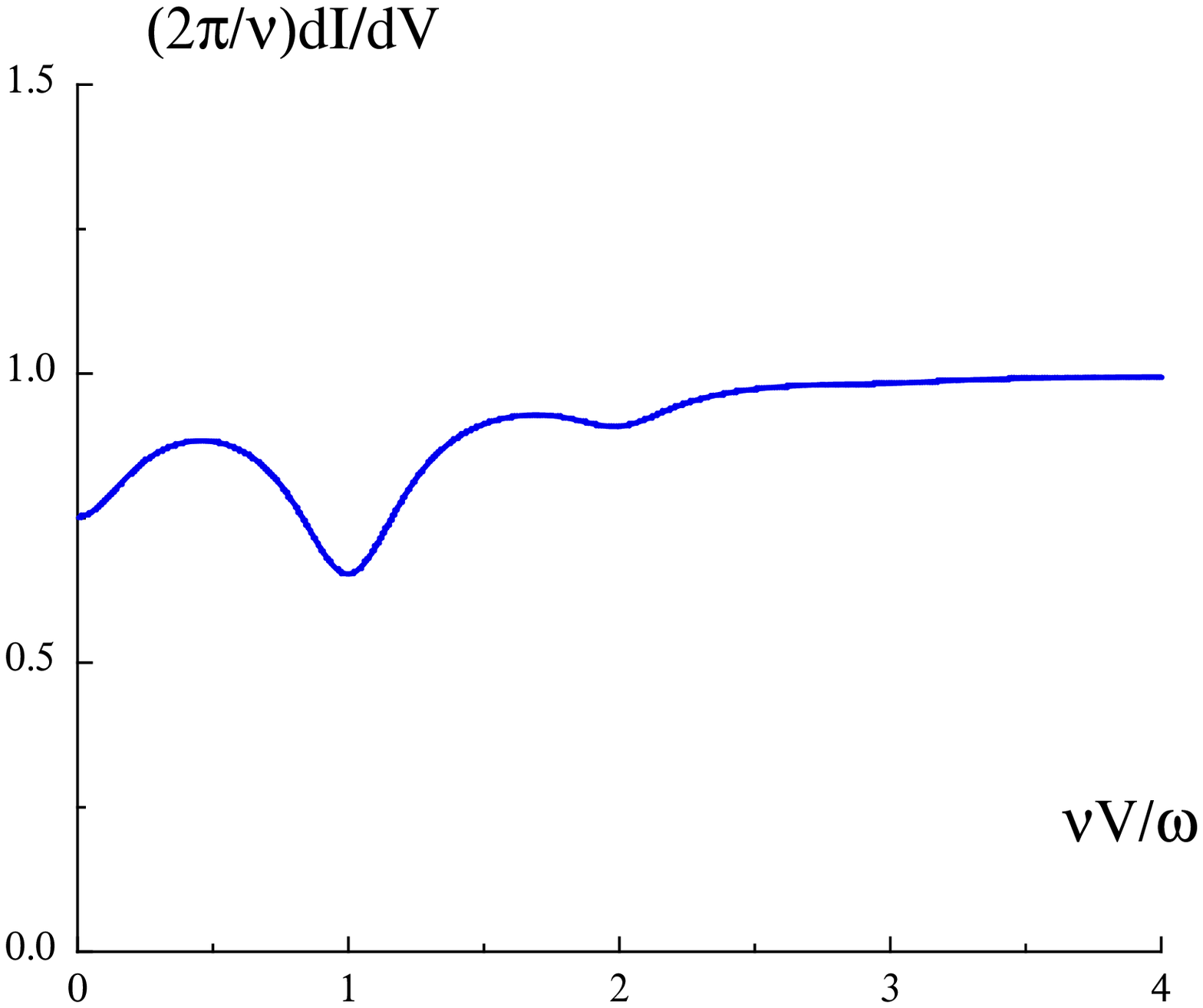}}
{\noindent FIG. 5:
Differential conductance with ac drive at $\nu=1/2$ 
obtained from (4.34), plotted versus voltage.
We have put $\nu T_B =\omega/4$, $\nu V_{ac}=1.6 \omega$. 
The minima correspond to 
smeared plateaus in the I-V curve, centered around $\nu V_{sd} = n\omega$.
}
\end{figure}

\begin{figure}
\epsfysize=3.2 truein
\epsfxsize=3 truein
\centerline{\epsfbox{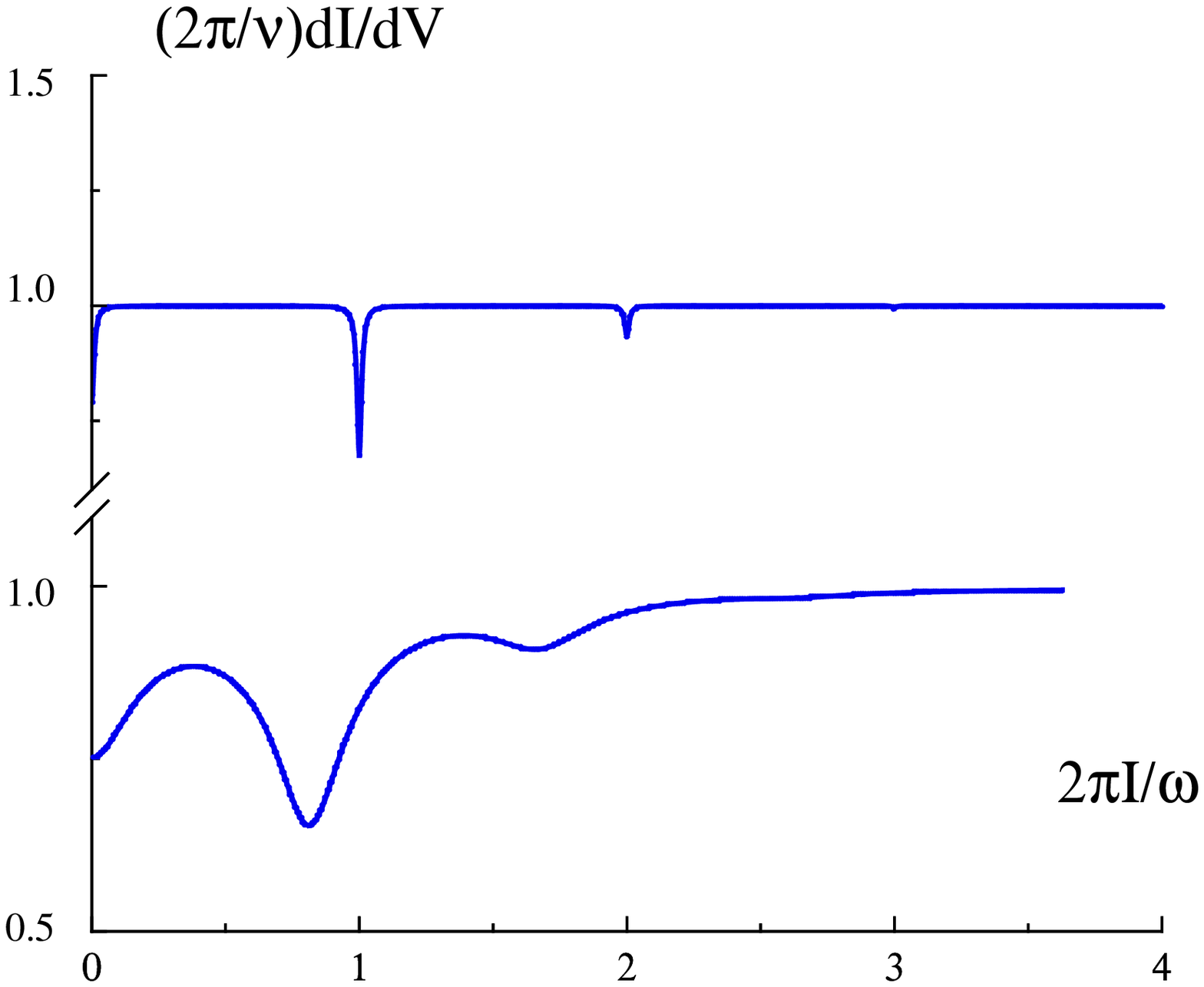}}
{\noindent FIG. 6:
Differiential conductance with ac drive at $\nu=1/2$ at two different
back scattering strength. In the lower part of the figure,
we plot the same differential
conductance  as in Figure 5 but versus current. 
As one can see that the smeared
plateaus are {\it not} centered at $I= ne\omega/2\pi$ due to the
finite offset. In the upper part, we put 
$\nu T_B =\omega/100$ and $\nu V_{ac}=1.6 \omega$.
In this weak backscattering limit,
the ``plateaus'' are centered at the quantized 
values, $I = ne\omega/2\pi$.
}
\end{figure}

It is also instructive to plot the differential conductance,
$dI/dV$, 
versus the current, as shown in Figure 6.
This shows clearly that the 
smeared ``plateaus" are {not} centered at
the quantized current values $I=ne \omega/2 \pi$.
This is due in part to the finite offset voltage at large bias,
mentioned above.  However, if we choose a smaller backscattering
strength, $T_B \rightarrow 0^+$, these ``plateaus" become centered
at the quantized current values, $I=ne \omega/2 \pi$,
as shown in Figure 6.

We next consider the I-V curve for general $\nu$.

\section{general $\nu$}

In the previous section we showed that
for both $\nu=1$ and $\nu=1/2$, the I-V curve with a.c. drive,
could be related to the I-V curve in the absence of any a.c. drive,
as a weighted sum over absorption and emission of quanta, see (4.34).
Here we make the conjecture that this relation holds in general, for
arbitrary $\nu$.
If correct, this conjecture allows us to use the
recent results of Fendley et. al.\cite{fendley}
to extract I-V curves with a.c. drive for {\it arbitrary} integer $\nu^{-1}$.
Before doing so, we
show that the conjecture does hold for general $\nu$
in the limit of weak backscattering.

To this end, consider calculating the I-V curve with a.c. drive
present, as a perturbation expansion in powers of the
backscattering amplitude $v$.
The leading non-vanishing correction appears
at order $v^2$.
This correction can be readily obtained
from the Keldysh action discussed in Sec III. 
The back scattered current operator follows from
(3.6) upon functional differentiation 
with respect to the gauge field $A(t)$,
\begin{equation}
\hat I_B = \nu v \sin (\theta + \nu A).
\end{equation}
Within the Keldysh approach, the average over this operator can be performed by
putting it into the forward path, so that
\begin{equation}
<I_B> = <\sin(\theta_+(t) + \nu A)>  ,
\end{equation}
where the average is taken with respect to the generating functional
(3.7).
Upon expanding the exponential $exp(-S_1)$ to second order
in $v$, one obtains
\begin{eqnarray}
&&\hspace{.2in}I_B = -\frac{i}2 \nu v^2 \int dt'\nonumber\\ 
&&\left< \sin[\theta_+(t)-\theta_-(t')+ \nu (A(t)-A(t'))] \right>_0,
\end{eqnarray}
where the subscript $0$ denotes an average
with respect to the quadratic action $S_0$ in (3.9).
Keeping only the constant time-independent piece,
it is then straightforward to show that,
\begin{equation}
<I_B>_{time} = \sum_n |c_n|^2 \; I_B^{dc}(\nu V_{sd} + n \omega),
\end{equation}
where $I_B^{dc}$ is the backscattered current to order $v^2$
without the a.c. drive.  Since the total current is
$I= \nu V_{sd} - I_B$, the sum rule $\sum_n |c_n|^2 =1$ 
can be used to verify the general conjecture (1.1). A similar
expression for the current under ac drive in small backscattering
regime was obtained by Wen before\cite{wen}.

The general conjecture has a very simple and physical interpretation.
In the absence of an a.c. drive,
Laughlin quasiparticles with fractional charge $\nu$
lose an energy $\nu V_{sd}$ when they backscatter from one edge to the other.
With a.c. drive, the quasiparticles absorb or emit
quanta of energy $\hbar \omega$ from the ac drive and jump into different energy
levels, $\nu V_{sd} \pm n \omega$, with probability $|c_n|^2$.
They are then backscattered by the point contact,
with reflection coefficient given by the energy dependent
S-matrix.  This contributes a backscattered current $I^{dc}(\nu V_{sd} \pm n
\omega)$.  The total backscattering follows by summing
over the number of absorbed quanta $n$, weighted
with the probability
$|c_n|^2$.

\begin{figure}
\epsfysize=3.5 truein
\epsfxsize=3 truein
\centerline{\epsfbox{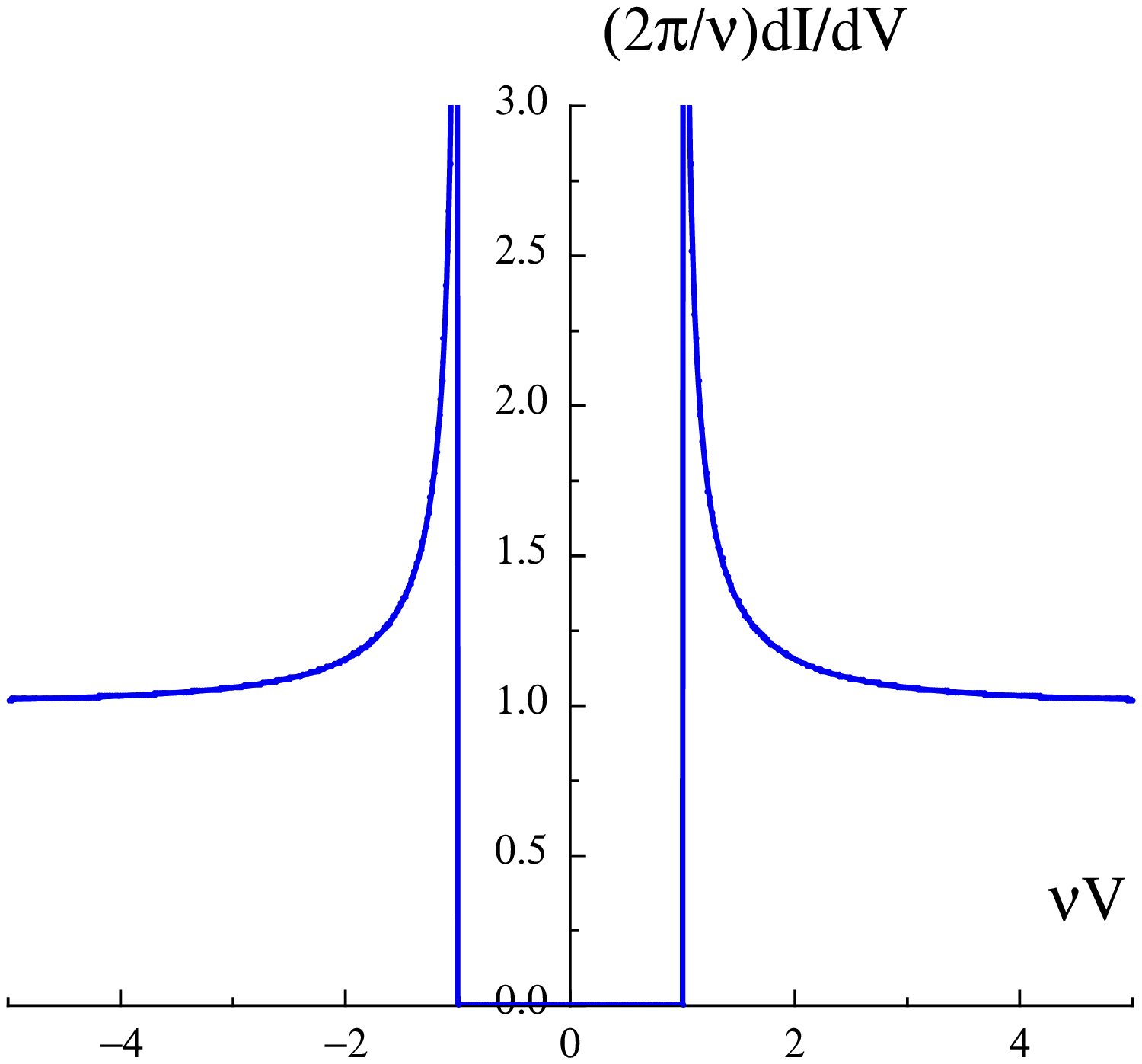}}
{\noindent FIG. 7:
Differiential conductance with no ac drive present
in the $\nu \rightarrow 0$ limit with $\nu T_B=1$ fixed,
as obtained from the exact solution of 
Fendley et. al..  This I-V curve is identical to that obtained
from the classical equation of motion (3.17).
Below a threshold voltage there is {\it no} electron tunnelling.
}
\end{figure}

In the absence of ac drive, the I-V curve can be extracted from the exact 
solution of Fendley et. al.. We can then use the general conjecture to
evaluate explicitly the I-V curve with ac drive for the Laughlin FQHE
states. It is interesting to compare the results in 
the $\nu \rightarrow 0$ limit with semi-classical model in Sec III.
Using the exact solutions of Fendley et. al.,
it is possible to take the $\nu \rightarrow 0$ limit.
If we keep $\nu V_{ac}, \nu V_{sd}, \nu T_B$
fixed as $\nu \rightarrow 0$, then the $I-V$ curve (with no ac drive),
reduces to that obtained from the classical equation of motion (3.17).
This solution is shown in  
Figure 7.
With ac drive present, the $I-V$ curves in the $\nu \rightarrow 0$ limit
can be extracted from our conjecture, and are shown in Figure 8.
Notice that these $I-V$ curves do {\it not} 
coincide with
the solutions of the classical equation
of motion (3.17), plotted in Figure 2.
Specifically, they do {\it not} exhibit flat mode-locked plateaus,
in contrast to Figure 2.
As discussed at the end of Section III,
the reason for this can be attributed to the presence of two
parallel quantum processes which facilitate electron transfer:
tunnelling under the barriers, and over the barrier motion after
quantized energy absorption from the ac field.
In the $\nu \rightarrow 0$ limit, only the first process is suppressed.
However, in the classical limit (3.17), both quantum processes
are absent.  The second quantum process is evidently responsible
for smearing the current plateaus, and spoiling the precise
quantization.  
	
\begin{figure}
\epsfysize=3.5 truein
\epsfxsize=3 truein
\centerline{\epsfbox{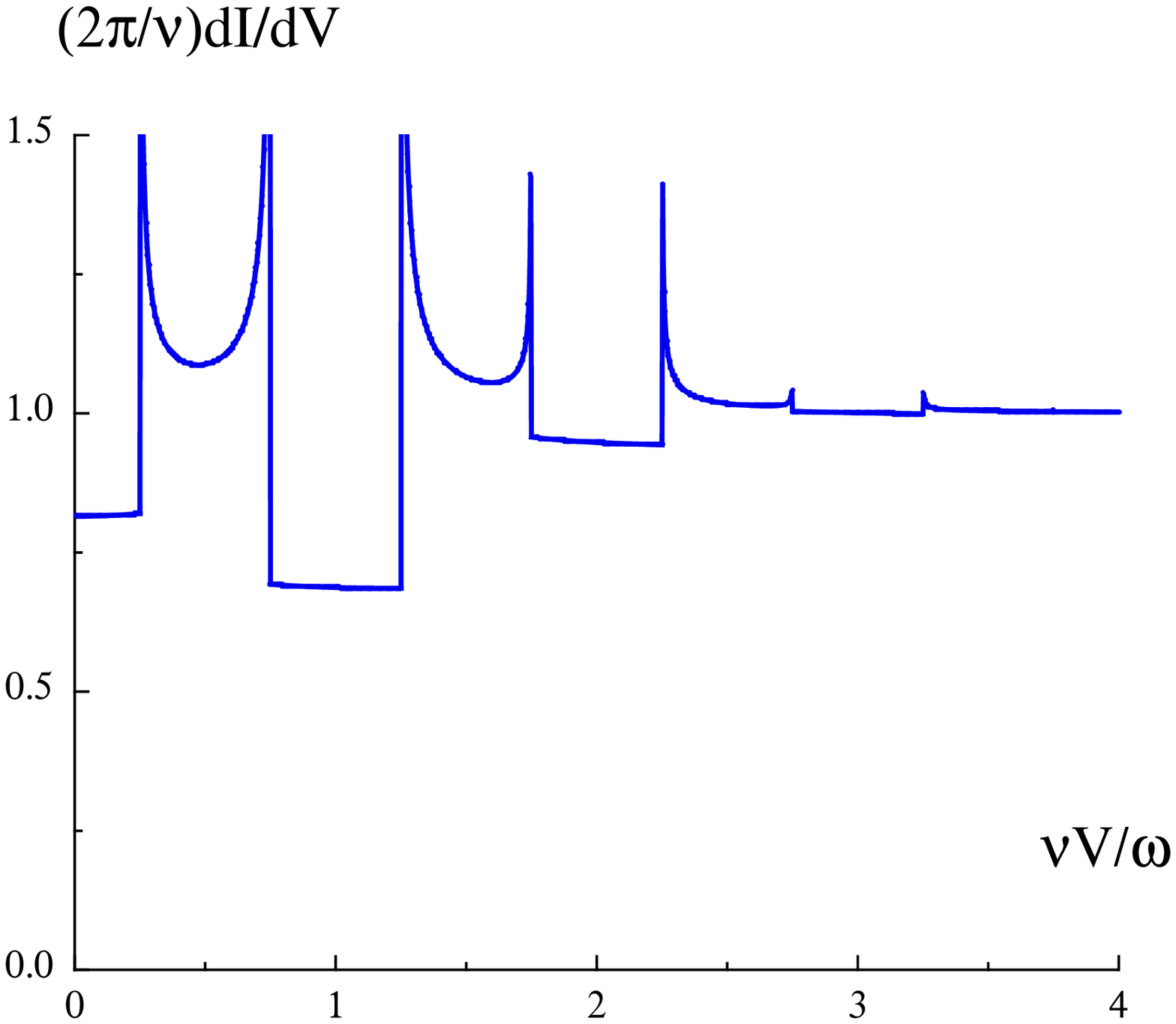}}
{\noindent FIG. 8:
Differiential conductance with ac drive present
obtained from the conjecture (1.1),
in the $\nu \rightarrow 0$ limit with $\nu T_B = \omega /4$ and 
$\nu V_{ac} = 1.6 \omega$ held fixed,  
plotted verus voltage.
In contrast to the classical solution from (3.17)
shown in Figure 2, the above current
``plateaus" are rounded by quantum fluctuations.
}
\end{figure}

\section{Discussion and conclusions}

Electron turnstile devices which transfer
electrons one-by-one have been made both in
metals\cite{geerligs,geerligs1,white} and in
semiconductors\cite{kouwenhoven,kouwenhoven1,nagamune}.
These devices consist of multiple tunnel junctions in series.
By selectively controlling the junction barrier heights a
single electron at a time can be transported through
the device.  By applying an ac drive field to the barriers,
with an appropriate phase relationship between successive junctions,
it is possible to transport one electron through the device for each period of the
ac drive.  This leads to a plateau in the current-voltage characteristics
at the quantized value $I = e \omega /2 \pi$, with $\omega$ the ac drive 
angular frequency.  These devices work well only at relatively
low frequencies (Mega Hertz).  At higher frequencies, the electrons
do not ``keep up" with the ac drive.

A key feature of the turnstile devices appears to be the multiple junction
geometry.  The electron being transported through the device
presumably suffers numerous inelastic scattering events
while on the islands between junctions.  These phase breaking
events destroy the electron coherence, effectively making the electron dynamics
classical and suppressing leakage from quantum tunnelling.
In this way quantum fluctuations do not destroy the mode
locking to the external ac drive.

In this paper, we have considered a point contact tunnel junction
between two quantum Hall fluids.  In contrast to the turnstile
devices, the junction has only one barrier.
The electron is assumed to tunnel coherently through the
barrier, only suffering inelastic collisions
in the ``contacts" (ie in the edge states) on either
side of the barrier.  
Moreover, we assume the transmission amplitude
through the barrier to be independent of energy.
Within a semi-classical approximation to this model,
there are robust quantized current plateaus in the presence
of an ac drive field, similar to
that seen in the turnstile devices.
However, inclusion of quantum fluctuation effects tends to smear
these plateaus.  Specifically,
for the IQHE
at filling $\nu=1$ where the edge state is a free fermion gas,
the I-V curve with ac drive is strictly Ohmic and featureless.
Mode locked plateaus in the current are completely
destroyed by quantum fluctuations.

In the FQHE the edge states are Luttinger liquids.
In this case, even though the bare tunnelling amplitude through
the point contact is energy independent, interaction effects
in the edge state ``leads" give an energy dependence
to the total tunnelling rate.  (The tunnelling rate
vanishes as a power law of energy, for energies close to the Fermi energy.)
In this case, with an ac drive present,
the $I-V$ curves do exhibit features, which can be identified
as mode locked current plateaus rounded by quantum fluctuations.
However, the ``plateaus" are {\it not} completely flat - 
the current varies upon sweeping the voltage.
In the limit of weak backscattering at the point contact,
the rounded plateaus are centered at
currents given by integer multiples of
$I = e \omega/ 2\pi$, as shown in Figure 6.

What are the prospects for using FQHE tunnel junctions
as a current to frequency standard?  
The prediction of rounded current plateaus
induced by the ac drive -
centered at quantized values $I = n e \omega/ 2\pi$ for
weak backscattering - is encouraging.  .  However, since the 
``plateaus" are {\it not} completely flat, the degree of current
quantization will necessarily be limited.  
Whether this fundamental limitation renders
the device useless, is difficult to assess.  In a practical device, the quantization
could presumably
be improved upon, by making multiple FQHE junctions, similar to the
semiconductor turnstiles.  In any event, it would be
fascinating to explore experimentally the
effects of an ac drive on quantum Hall point contacts.

We thank A. Ludwig, S.M. Girvin and Paul Fendley for
clarifying conversations.  We are grateful to the National
Science Foundation for support under grants PHY94--07194, DMR--9400142
and DMR-9528578.

\end{multicols}
\end{document}